\documentclass[twocolumn,showpacs]{revtex4}
\usepackage[dvips]{color}
\usepackage{graphicx}
\usepackage{dcolumn}

\def\lsim{\mathrel{\raise2pt\hbox to 8pt{\raise -5pt\hbox{$\sim$}\hss{$<$}}}}
\def\bmath#1{\mbox{\boldmath $#1$}}

\begin{document}
\title{Isospin Asymmetry in the Pseudospin Dynamical Symmetry}
\author{P. Alberto, M. Fiolhais}
\affiliation{Departamento de F\'\i sica and Centro de F\'\i sica
Computacional,
Universidade de Coimbra, P-3004-516 Coimbra, Portugal}
\author{M. Malheiro, A. Delfino}
\affiliation{Instituto de F\'\i sica, Universidade Federal Fluminense, 
24210-340 Niter\'oi, Brazil}
\author{M. Chiapparini}
\affiliation{Departamento de F\'\i sica Te\'orica, Universidade do Estado
 do Rio de Janeiro,
20550-900 Rio de Janeiro, Brazil}
\date{\today}
\begin{abstract} 
\noindent

Pseudospin symmetry in nuclei is investigated considering the Dirac
equation with a Lorentz structured Woods-Saxon potential.  
The isospin correlation of the energy splittings of pseudospin partners
with the nuclear potential parameters is studied. We show that, in 
an isotopic chain, the pseudospin symmetry is better realized for neutrons 
than for protons. This behavior comes from balance effects among the 
central nuclear potential parameters.  In general, we found an isospin
asymmetry 
of the nuclear  pseudospin interaction,  opposed to the nuclear spin-orbit
interaction 
 which is quasi isospin symmetric.  
\end{abstract}
\maketitle

In some heavy nuclei a quasi-degeneracy is observed 
between single-nucleon states with quantum numbers
($n$, $\ell$, $j= \ell + 1/2$) and
($n-1$, $\ell+2$, $j= \ell$ + 3/2)
where $n$, $\ell$, and
$j$ are the radial, the orbital, and the total angular momentum
quantum numbers, respectively.
This doublet structure is better expressed
using a ``pseudo'' orbital angular momentum quantum number,
$\tilde{\ell} = \ell + 1$, and a ``pseudo'' spin quantum number,
$\tilde s = 1/2$.
For example, for $[n s_{1/2},(n-1) d_{3/2}]$ one has $\tilde{\ell}= 1$,
for $[n p_{3/2},(n-1) f_{5/2}]$ one has $\tilde{\ell}= 2$, etc.
Exact pseudospin symmetry means  degeneracy of
doublets whose angular momentum quantum numbers are
$j = \tilde{\ell}\ \pm \tilde s$. This
symmetry in nuclei was first reported about 30 years ago
  \cite{kth}, but only recently  has its  origin
become a topic of intense theoretical research.

In recent papers \cite{bahri,mosk,gino,arima,meng0}
possible underlying mechanisms to generate such
symmetry have been discussed.
We briefly review the main points of these studies.

 Blokhin {\em et al.} \cite {bahri}
performed a helicity unitary transformation in a nonrelativistic
single-particle Hamiltonian. They showed that the transformed radial
wave functions have a different asymptotic behavior,
implying that the helicity
transformed mean field acquires a more diffuse surface.
Application of the helicity operator to the 
nonrelativistic single-particle
wave function maps the normal state
$(l,s)$ onto the ``pseudo" state $(\tilde l , \tilde s )$, while
keeping all other global symmetries \cite{bahri}.  The same kind of unitary
transformation was also considered earlier by Bahri {\it et al.}~\cite{mosk}
to discuss the pseudospin
symmetry in the nonrelativistic harmonic oscillator.   
They showed that a particular condition between the
coefficients of spin-orbit and orbit-orbit terms, required to have a 
pseudospin symmetry in that non-relativistic single particle Hamiltonian, 
was consistent with relativistic mean-field (RMF) estimates.

Ginocchio \cite {gino}, for the first time, identified the pseudo\-spin
 symme\-try as a symme\-try of the Dirac Hamiltonian. 
He pointed out that the
pseudo-orbital angular momentum is just the orbital angular momentum of
the lower component of the Dirac spinor. Thus, the pseudospin
symmetry started to be regarded and understood in a relativistic
way. He also showed that the pseudospin symmetry would be
exact if the attractive scalar, $S$, and the repulsive vector, $V$, 
components of a Lorentz structured potential were equal in 
magnitude: $S+V=0$. 
Under this condition,  
the pseudospin symmetry was identified as a SU(2) symmetry of the 
Dirac Hamiltonian \cite{levi}.

In RMF models, 
%often referred to as Quantum Hadrodynamics, the
%nuclear saturation mechanism is explained as 
%a consequence of
nuclear saturation is explained
by a cancellation between a large scalar ($S$) and a large
vector ($V$) fields \cite {furn0}.  Typical values for
these fields in heavy nuclei are of the order of a few hundred MeV
(with opposite signs), their sum providing a binding potential of about
$60$~MeV at the nucleus center.  Therefore, a natural claim of Ginocchio
was to
regard the quasi-degenerate pseudospin doublets in nuclei as arising
from the near equality in magnitude of the attractive scalar, $S$, and
the repulsive vector, $V$, relativistic mean-fields, $S \sim - V$, in
which the nucleons move. 
When the doublets are degenerate, the shape of the lower components 
of the Dirac spinor for the two states in the doublet is the
 same~\cite{gino2}. 

More recently, Meng {\em et al.} \cite{arima,meng0} showed that
pseudo\-spin symmetry is exact when $d \Sigma/dr=0$ where $\Sigma=S+V$.
They also related the onset of the pseudospin symmetry to a
competition between the centrifugal barrier and the
pseudospin-orbit potential.  

Actually, since in nuclei $S(r),\,V(r)\rightarrow 0$ 
%at large distances,
%the conditions $\Sigma=S+V=0$
%or
when $r\to \infty$,
the conditions $\Sigma=0$
and $d \Sigma/dr=0$ are equivalent.
However, these conditions cannot be realized in nuclei, because they imply
that no bound states
exist \cite{gino}.  Nevertheless, as we will show, there is a correlation
between the pseudospin splitting and the depth of $\Sigma$, its
surface diffuseness and its radius.  Depending on the actual choice of these
parameters, fitted to describe a nucleus, the pseudospin symmetry may show up.
Thus, one may argue that pseudospin symmetry is a dynamical symmetry in
nuclei, in the sense of Arima's definition of a dynamical
symmetry~\cite{arima1}:  (i) a symmetry of the Hamiltonian which is not
geometrical in nature; or (ii) a ordered breaking symmetry from
dynamical reasons. This is consistent with the findings of
Ref.~\cite{mosk},  associating the pseudospin
symmetry with a particular relation between the coefficients
of spin-orbit and orbit-orbit terms in RMF models.

To establish that correlation, we perform a model calculation
with a Lorentz
structured potential of Woods-Saxon type in the Dirac
equation.  The scalar and vector components of this potential are the
mean-field central nuclear potentials.  The form of the potential is
\begin{equation}
U(r)={U_0\over {1+\exp [(r-R)/a]}}\ ,
\label{WSaxon}
\end{equation}
where $U(r)$ stands for either the vector or the scalar potential.
Although this is not a full self-consistent relativistic potential,
the use of (\ref{WSaxon}) as a nuclear mean-field is
realistic enough to
be applied to nuclei and enables us to study
the splittings of pseudospin partners not only as a function of the
mentioned depth, $U_0$, but also of the diffuseness, $a$, and the radius, $R$.

%The surface diffuseness is usually expected to depend strongly on
% the nucleon separation energy.  In the case of isotopes it is known that the
%central depth and the surface diffuseness changes sensibly
%\cite{nikolaus,meng}.  In the extreme situations of halo nuclei, very small
%neutron binding energies may occur 
%\cite{tanihata}.  Consequently, one may have drastic different choices on
%the parameters in order to fit the data. We have considered parameters
%in our potentials
%which reproduce the experimental single-particle spectra of neutrons in
%$^{208}$Pb.

In nuclear relativistic mean-field theory, each nucleon is described by a 
 Dirac Hamiltonian of a particle of
mass $m$ in an external scalar, $S$, and vector,
$V$, potentials:
\begin{eqnarray}
H = \mbox{\boldmath $\alpha\cdot p$}
+ \beta (m + S) + V  ~,
\label {dirac}
\end{eqnarray}
where \mbox{\boldmath $\alpha$} and $\beta $ are the usual Dirac
matrices.  The Dirac Hamiltonian is invariant under an SU(2)
transformation for two cases:  $S$ = $V$ and $S = - V$
\cite{smith,bell}. 

 We define 
$\Delta=V-S$ and denote the upper and lower components of the
Dirac spinor by $\Psi_{\pm}={1\pm\beta\over 2}\Psi$.
Assuming that $S$ and $V$ are radial functions, the Dirac equation can be
decoupled
into two equations for the lower and upper components, respectively: 
\begin{eqnarray}
\nabla^2 F_i+{\Sigma'\over E-m-\Sigma}\bigg(F'_i+
{1-\kappa_i\over r}F_i\bigg)+\nonumber\\
(E+m-\Delta)(E-m-\Sigma)F_i=0
\label {lower}
\\
\nabla^2 G_i+{\Delta'\over E+m-\Delta}\bigg(G'_i+
{1+\kappa_i\over r}G_i\bigg)+\nonumber\\
(E+m-\Delta)(E-m-\Sigma)G_i=0 \, ,
\label{upper}
\end{eqnarray}
where the primes 
denote derivatives with respect to $r$. The spinors
$\Psi_\pm$ have been factorized in radial and angular parts:
$\Psi_+={\rm i}\,G_i(r)\,\Phi^+_i(\theta,\phi)$ and
$\Psi_-=-F_i(r)\,\Phi^-_i(\theta,\phi)$,
 with $i$ standing for the quantum numbers of the single particle state.
The property
$\bmath{\sigma}\cdot\bmath{L}\Phi^\pm_i=-(1\pm\kappa_i)\Phi^\pm_i$
was used, $\bmath{\sigma}$ being the Pauli matrices.

Before we present our results, let us briefly discuss how the pseudospin
symmetry gets broken. 
%The SU(2) generator of the pseudospin symmetry is given
%by \cite{levi,smith,bell,dudek}
%\begin{equation}
%S_i
%=\pmatrix{\tilde s_i& 0\cr 0&s_i\cr}\ ,
%\end{equation}
%where $\tilde s_i={\bmath\sigma\cdot\bmath p\over p} s_i
%{\bmath\sigma\cdot\bmath p\over p}={2\bmath s\cdot\bmath p\over p^2}
%p_i-s_i$ and $s_i={\sigma_i/2}$.
%
%The commutator of this operator with the 
%Hamiltonian (\ref{dirac}) is given by 
%\begin{equation}
%[H,S_i]=\pmatrix{[\Sigma,\tilde s_i]&0\cr 0&0\cr}\, .
%\label {comut}
%\end{equation}
The commutator of the SU(2) generators of pseudospin symmetry
with the Hamiltonian (\ref{dirac}) is given by 
\cite{levi,smith,bell,dudek}
\begin{equation}
[H,S_i]=\pmatrix{[\Sigma,\tilde s_i]&0\cr 0&0\cr}\ ,
\label {comut}
\end{equation}
where $\tilde s_i={\bmath\sigma\cdot\bmath p\over p} s_i
{\bmath\sigma\cdot\bmath p\over p}={2\bmath s\cdot\bmath p\over p^2}
p_i-s_i$ and $s_i={\sigma_i/2}$.
The breaking of the pseudospin symmetry can thus be related to the
commutator $[\Sigma,\tilde s_i]$. 
 Requiring $[\Sigma,\tilde s_i]=0$ is equivalent to
the previous condition $d \Sigma/dr=0$, when $\Sigma$
is a radial function.

Now we turn to the presentation of our results obtained using
the Woods-Saxon potentials (\ref{WSaxon}) in eqs. (\ref{lower}) 
and (\ref{upper}). There are altogether six parameters for 
$\Sigma$ and $\Delta$, namely the central depths, $\Sigma_0$
and $\Delta_0$, two radii and two diffuseness parameters. 
We observed that the pseudospin splitting is not sensitive to
$R$ and $a$ of the $\Delta$ potential, and, accordingly, set
the same radius, $R$, and surface diffuseness, $a$, for both potentials. 
We first fitted these
parameters to the neutron spectra of $^{208}$Pb, obtaining a good 
agreement with the results of 
Ref. \cite{furn1} for the same set of pseudospin doublets:
  $(1i_{11/2},\ 2g_{9/2})$, $(2f_{5/2},\ 3p_{3/2})$ and 
$(1h_{9/2},\ 2f_{7/2})$.

 Keeping
$\Sigma_{0}$ and  $\Delta_{0}$ fixed, we varied $a$ and  $R$ in order
 to see
how the energy splittings of the pseudospin doublets change with the
surface diffuseness and the radius.  This dependence is presented 
in Fig.~1. As $a$
increases, the splittings of the pseudospin doublets decrease. 
On the contrary, as $R$ increases the pseudospin splittings increase.  
The same figure shows that the splittings are more dependent
on  $R$ than on $a$.  
If $a$ increases and/or $R$ decreases enough, the pseudospin
doublet partners cross
each other, inverting the sign of the energy splitting, and are driven
apart if $a$ ($R$) further increases (decreases).  This inversion
of pseudospin partner splittings as $a$ increases, $E_{n-1,\tilde l +
1/2} < \,\,E_{n,\tilde l - 1/2}$ changing to $E_{n-1,\tilde l + 1/2} >
\,\,E_{n,\tilde l - 1/2}$, is observed experimentally and was also found
in~\cite{meng0,ring,marcos}. 
\begin{figure}[t]
\begin{center}
%\vspace*{-.2cm}
\includegraphics[clip=on,width=6.3cm,angle=-90]{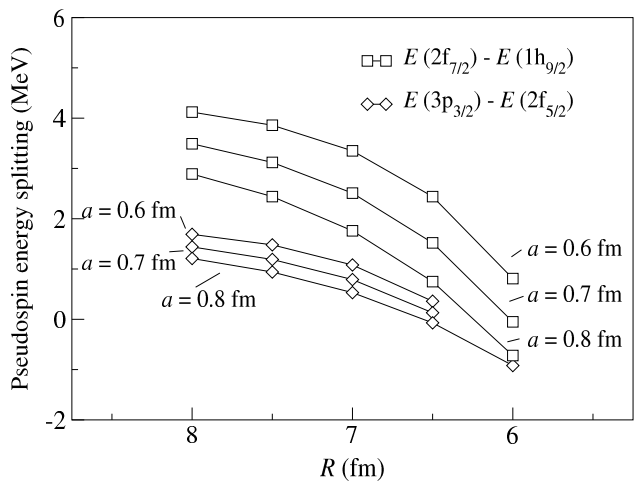}
\end{center}
\vspace*{-.3cm}
\caption[Figure 1]{Pseudospin energy splittings for the neutron
 pseudospin partners 
 $(2f_{5/2}, 3p_{3/2})$ and $(1h_{9/2}, 2f_{7/2})$ in $^{208}$Pb as a
function of
$R$ for several values of $a$.}
\vspace*{-.2cm}
\end{figure}

\begin{figure}[hbt]
\begin{center}
%\vspace*{-.2cm}
\includegraphics[clip=on,width=6.3cm,angle=-90]{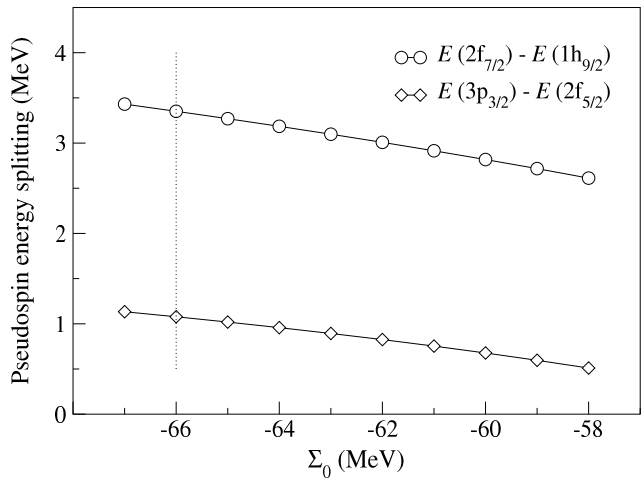}
\end{center}
\vspace*{-.3cm}
\caption[Figure 2]{Pseudospin energy splitting in $^{208}$Pb as a
function of the depth of the $\Sigma$ potential for the same
neutron pseudospin partners (similar to Fig.~1). The vertical
line corresponds to the fitted Woods-Saxon parameters for
$^{208}$Pb: $\Sigma_0 = -66 $ MeV, $\Delta_0=650$ MeV, $R = 7$ fm
 and $a = 0.6$ fm.}
\vspace*{-.2cm}
\end{figure}

We now keep $a$, $R$ and $\Delta_0$ fixed to study the sensitiveness
of the pseudospin doublets with $\Sigma_0$. The results are
presented in Fig.~2 and one observes that, as
$|\Sigma_0|$ decreases, the splitting also decreases.  This is in
accordance with Ginocchio predictions for pseudospin symmetry breaking
due to the finiteness of the $\Sigma$ mean field. We also found that,
for deeper levels, an inversion of pseudospin energy splitting occurs
for sufficiently low $|\Sigma_0|$. 
Comparing Figs.~1 and 2, we see that the
splittings of pseudospin partners are more sensitive to the 
radius and diffuseness than to the central depth of the potential.

We note that the radius dependence is
relevant for comparing different nuclei, especially certain    
isotopes \cite{meng}. However, we found an interesting correlation between
$|\Sigma_0|$ and $R$. Varying $|\Sigma_0|$ and $R$ but 
keeping $a$ and the product $|\Sigma_0|R^{2}$ fixed, the pseudospin 
doublet splittings remain almost constant.
% as shown in Fig.~3.
Therefore, only two parameters, out of
($\Sigma_{0},\, R ,\,a$), are free.
%
%\begin{figure}[ht]
%\begin{center}
%\includegraphics[clip=on,width=6.3cm,angle=-90]{fig3.eps}
%\end{center}
%\vskip-.2cm
%\caption[Figure 3]{Pseudospin energy splitting in $^{208}$Pb as a
%function of $R$ keeping 
%$a$ and the product $|\Sigma_0|R^{2}$ fixed for 
% the same
%neutron pseudospin partners (similar to Fig.~1).}
%\end{figure}

The variation of the other free parameter, $\Delta_0$,
does not qualitatively change the splittings. Since the $\Delta$
potential is related to the effective mass of the nucleons, the basic
effect is to slightly change the nucleon separation energies, and
especially the spin-orbit coupling, as discussed below. 
We have also performed the calculation for calcium isotopes, and found
the same dependence of the pseudospin splitting with $R$, 
$a$ and $\Sigma_0$.

We apply now the systematics to some nuclei studied in the literature. 
 Recently, the pseudospin symmetry in
Zr and Sn isotopes was investigated as a function of the number of
nucleons \cite{meng0}.  The form of the $\Sigma$ potential in dependence
of the radial distance for such nuclei, as $A$ increases, is given in
Ref.  \cite{meng}, starting with $^{100}$Sn and going up to
$^{170}$Sn. In~\cite{meng} it is shown that 
the central neutron depth, $\Sigma_0$, of Sn isotopes varies from about 
$-65$~MeV to
$-54$~MeV, $R$ from 5.6~fm to 6.8~fm and the surface diffuseness increases.
For the proton, the central depth, $\Sigma_0$, varies from $-50$~MeV
 to $-60$~MeV, $R$ from 5~fm to 6.5~fm, while  
 the surface diffuseness also increases, now slightly more than in 
the neutron case.  
Using (\ref{WSaxon}), good fits to the neutron central 
potentials of the Sn isotopes were obtained, with $\Sigma_0$ from $-68.6$~MeV 
to $-55.7$~MeV, $R$ from 5.6~fm to 6.7~fm and  $a$ from 0.6~fm to 0.76~fm, in 
agreement with~Ref.~\cite{meng}. 
 
This information on $\Sigma$ for Sn isotopes allows us to analyze the
behavior for
neutron pseudospin doublets going from $A=100$ to $A=170$.  As $A$
increases, the central depth $|\Sigma_0|$ decreases and the
surface diffuseness increases, both effects favoring the pseudospin
symmetry, as shown in Figs.~1 and 2. However, the radius increase 
with $A$ can partially offset those effects. Since the values 
of $|\Sigma_0|R^{2}$ are roughly constant for neutrons \cite{meng}, 
the correlation between these two parameters,
%, shown in Fig.~3,
mentioned above, implies that the effects of increasing $R$ and 
decreasing $|\Sigma_0|$ in the neutron central potential, when $A$ increases,
balance each other. Thus, the dominant effect comes from 
the increasing $\,a\,$, slightly favoring the pseudospin symmetry.   

We extended our analysis to the proton spectra.  For this case, as
mentioned above, 
 $|\Sigma_0|$ increases as $A$ increases
for Sn isotopes \cite{meng}, therefore disfavoring pseudospin symmetry.
 Moreover,
$R$ increases, also disfavoring the pseudospin splitting. Hence, at
least for the proton spectra, and in opposition to the neutron spectra,
 we do not have any balance effect coming from $\Sigma_{0}$ and 
$R$ dependences.
Thus, we have a competition between the effects of surface diffuseness 
(favoring the symmetry), 
and of depth and radius of the central potential (both disfavoring
the symmetry). From our systematics described before we expect that
effects coming from the depth and radius of the central potential 
override the surface diffuseness dependence. Therefore,  
although the proton spectra along
the Sn isotopic chain was not presented in Ref. \cite{meng0}, 
our systematics predicts that 
the splitting increases for the proton pseudospin partners.
The different behavior of the splitting for protons and neutrons can  be
inferred by looking at the value of $|\Sigma_0|R^{2}$ for the two cases as $A$
changes: the change is considerably higher for protons.

 The reason for the different behavior of the central potential depth
for neutrons and protons as a function of $A$, lies on
the $\rho$ meson interaction, which is repulsive for neutrons and
attractive for protons, as explained by mean-field model
calculations \cite{furn1}.  The inclusion of this interaction, which is
important in asymmetric nuclei, changes the vector part of the
nuclear potential $V$ to
\begin{equation}
V = V_{\omega}\pm {g_{\rho}\over {2}} \rho_0\, ,
\label{V_rho}
\end{equation}
with + and $-$ signs for protons and neutrons respectively; $\rho_0$ is
the time component of the $\rho$ field, which is proportional to the
number of protons minus the number of neutrons, and $V_{\omega}$ comes
from the vector-isoscalar $\omega$ meson. The increase of the parameters
$R$ and $a$ with $A$ can be traced back to the known nuclear
radius $A^{1/3}$ dependence and to the excess of neutrons
on the surface (the neutron skin effect). Hence, we may conclude that,
for a given nucleus, the parameters $\Sigma_0$, $a$ and $R$ for
protons and neutrons are different. Then, an isospin asymmetry
in the pseudospin interaction is expected to take place in agreement with the 
analysis of the mean-field nuclear parameters for protons and neutrons done
in 
~\cite{mosk}. 
In particular, since $\rho_0$ is negative for heavy nuclei, the vector
potential
(\ref{V_rho}) is bigger (and, thus, $|\Sigma|$ smaller) for neutrons
than for protons. From our previous analysis, for a neutron rich nucleus, 
the pseudospin symmetry for neutron spectrum is favored, in agreement
 with the results presented by Lalazissis {\em et al.}~\cite{ring}.

The systematics discussed here seems to be quite general and a comment
on how
it affects the spin-orbit splitting is pertinent.  Variations of central
depth and surface diffuseness of the nuclear potential do not change as
much spin-orbit splittings as they do for the pseudospin-orbit splittings.
The reason is that to change substantially spin-orbit splitting one
needs a significant relativistic content, {\em i.e.} a large lower
component in
the Dirac spinor \cite{palberto}.  The spin-orbit splitting is
completely correlated with the nucleon effective mass \cite{furn2}.
 From Eq.~(\ref{upper}), we see that it will
depend strongly on $\Delta$.  In RMF models 
this potential carries a quite
large scale when compared with $\Sigma$ ($\Delta_0$ is around 650-750 MeV,
while $|\Sigma_0|$ is around 50-70 MeV).
Therefore, the $\rho$ meson potential, $V_{\rho}$, in the range 
4-8 MeV, becomes irrelevant compared to $\Delta$. This was
verified by recent numerical calculations \cite{lal,meng,chiappa},
showing that the difference between the values of $\Delta$  
for protons and neutrons (and, therefore, between the corresponding
values of the spin-orbit term) is very small. This explains why the
spin-orbit interaction 
is roughly isospin symmetric. 

On the contrary, the pseudospin splitting can change when there is
little relativistic content, {\em i.e.}~a small lower component.  This may
explain why the nonrelativistic treatment of Ref.~\cite{bahri} to
analyze the origin of the pseudospin symmetry 
works well in explaining the small pseudospin
splitting. The pseudospin-orbit interaction depends on $\Sigma$, as
one can see from Eq.~(\ref{lower}). Since $V_{\rho}$ cannot be
neglected compared to $\Sigma$, the different sign contribution of
$V_{\rho}$ in (\ref{V_rho}) for protons and neutrons has a relevant
effect on $\Sigma$, leading to the isospin asymmetry  
in the pseudospin interaction.

The systematics observed in our model calculation helps us
to understand the origin of the pseudospin symmetry and explains
the quasi-degeneracy of single-particle states occurring in the spectra of some
finite nuclei. The new radioactive nuclear beam facilities will provide more 
data to which these results can be applied.

%\vspace*{-0.4cm}
\begin{acknowledgments}
We acknowledge financial support from FCT (POCTI), Portugal, and
from CNPq/ICCTI Brazilian-Portuguese scientific exchange program.
\end{acknowledgments}

%\vspace*{0.1cm}

\end{document}